# Adaptive Decision-Making for Autonomous Vehicles: A Learning-Enhanced Game-Theoretic Approach in Interactive Environments

Heye Huang, Jinxin Liu, Shiyue Zhao, Boqi Li*, and Jianqiang Wang*

*Abstract*—This paper proposes an adaptive behavioral decision-making method for autonomous vehicles (AVs) focusing on complex merging scenarios. Leveraging principles from non-cooperative game theory, we develop a vehicle interaction behavior model that defines key traffic elements and integrates a multifactorial reward function. Maximum entropy inverse reinforcement learning (IRL) is employed for behavior model parameter optimization. Optimal matching parameters can be obtained using the interaction behavior feature vector and the behavior probabilities output by the vehicle interaction model. Further, a behavioral decision-making method adapted to dynamic environments is proposed. By establishing a mapping model between multiple environmental variables and model parameters, it enables parameters online learning and recognition, and achieves to output interactive behavior probabilities of AVs. Quantitative analysis employing naturalistic driving datasets (highD and exiD) and real-vehicle test data validates the model's high consistency with human decision-making. In 188 tested interaction scenarios, the average human-like similarity rate is 81.73%, with a notable 83.12% in the highD dataset. Furthermore, in 145 dynamic interactions, the method matches human decisions at 77.12%, with 6913 consistence instances. Moreover, in real-vehicle tests, a 72.73% similarity with 0% safety violations are obtained. Results demonstrate the effectiveness of our proposed method in enabling AVs to make informed adaptive behavior decisions in interactive environments.

## I. Introduction

### A. Motivation

THE key objective of behavior decision-making for autonomous vehicles (AVs) is to enable them with the decision-making abilities of experienced human drivers [1]–[3]. This involves satisfying the comprehensive performance such as the safety, efficiency, and comfort of the AV, while ensuring the safety of other traffic participants. Meanwhile, the abilities include enabling AVs to handle uncertainties in dynamic environments effectively. However, the enhancement of AV decision-making abilities is significantly challenged by the stochastic interactions in complex, dynamic environments.

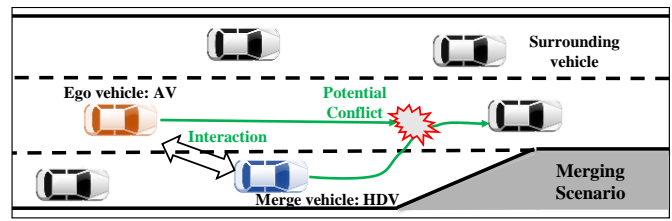

**Fig. 1.** Example of a dynamic interactive scenario.

As shown in **Fig. 1**, when faced with a scenario where the surrounding vehicles have uncertain motion status and random interaction behaviors, it is more challenging for AVs to achieve collision avoidance and rational driving. Moreover, in dynamic interaction scenarios, the interactive conflicts are a temporally extended process with uncertain interaction duration and unpredictable objects, which further complicates the task of rational interactive decision-making for AVs.

Therefore, how to achieve rational and safe behavioral decision-making with human-logic that adeptly responds to dynamic environmental changes emerges as a pivotal challenge. Technologies for AVs interactive decision-making are expected to address the inherent behavioral uncertainties and ensure transferability across various scenarios [4], [5]. To address these necessities, this paper introduces a novel method for modeling vehicle interactions and adaptive behavior decision-making in dynamic traffic. It simultaneously addresses the issues of poor adaptability of interaction models in uncertain environments and the difficulty of optimizing multiple parameters in decision models. Specifically, by employing offline modeling of vehicle interactions and a mapping model between environmental variables and model parameters, the AV can extract key environmental features in real time. It then identifies the optimal matching parameters for the interaction model, thereby enabling adaptive interactive decision-making in various traffic situations.

### B. Related Work

Existing methods for AV interactive decision-making can be categorized into four types [6], [7]: learning-based methods; probabilistic inference-based; potential field-based and game theory-based.

**Learning-based approaches.** In recent years, advanced deep neural networks [8], [9] like long short-term memory (LSTM), Social-LSTM [10], and graph neural networks (GNN) [11] are

This research was supported by National Natural Science Foundation of China, Science Fund for Creative Research Groups (52221005) and National Natural Science Foundation of China, the Key Project (52131201). Corresponding author: B. Li and J. Wang.

Heye Huang, Jinxin. Liu, Shiyue Zhao and Jianqiang Wang are with the School of Vehicle and Mobility, Tsinghua University, Beijing, 100084, China. (e-mail: hhy18@mails.tsinghua.edu.cn, jinxin_liu@foxmail.com, Stary132@163.com, wjqlws@tsinghua.edu.cn).

Boqi Li is with the Department of Civil and Environmental Engineering, University of Michigan, Ann Arbor, MI 48105, USA. (boqili@umich.edu)

employed for comprehensive modeling of vehicle data and environmental variables. Reinforcement learning (RL) techniques [12], [13] are further widely employed to generate behavioral decisions by considering the AVs as homogeneous intelligent agents within the traffic. Shi et al. developed learning-based models for agents' behavior control, these works prove the importance of adaptive decision-making, while ignore the dynamic interactions between multi-agents. Similarly, Tomizuka et al. [14] present a multi-agent RL approach that facilitates strategic right-of-way negotiations between AVs and drivers in merging scenarios. The inverse reinforcement learning (IRL) algorithm can deduce significant parameters within the reward function based on expert demonstrations or actions. It facilitates the understanding of the interplay between actions or strategies and reward function parameters. Consequently, it is frequently employed for optimizing the parameters of decision-making models. Although learning-based models are effective in handling simple interactive scenarios, such as IRL's effective realization of behavioral probability distributions, it is challenging to for GNN, RL based methods to interpret the interaction process as the complexity of driving situations increases. Specifically, they struggle to incorporate prior driving experiences or traffic regulations as reasonable constraints and generally ignore the impact of individual driving preferences on interactive behavior, limiting the widespread application.

**Probabilistic inference-based approaches.** These approaches can be categorized into two types. The first leverages probabilistic graphical models for interaction modeling between vehicles, outputting the probabilities of different interactive behaviors. The second formulates the interactive behavior decision problem as a Markov decision process [15], [16], using probabilistic expressions and mathematical optimization techniques to determine the optimal sequence of actions. A Bayesian driver agent model is introduced [17], by extracting traffic features and feeding them into a dynamic Bayesian network (DBN) for decision-making, thus facilitating both scene understanding and decision inference. Hubmann et al. [18] frame the problem as a partially observable Markov decision process (POMDP), using other vehicles as hidden variables to ascertain the optimal acceleration strategy. However, these probabilistic models sometimes struggle with establishing a comprehensive model structure, and suffer from slow optimization and overly conservative decision-making. Thereby posing challenges to their effective implementation in real-world dynamic scenarios.

**Potential field-based approaches.** They have gained widespread applicability, as employed in various studies such as automotive systems and robotics methods [19], [20]. Notable examples include driving safety fields and social force models [21], [22]. Vehicle interaction models based on the potential field employ this concept to define relational functions among traffic participants. These functions are used to characterize the interaction between the driver, vehicle, and road environment, and can comprehensively achieve vehicle behavior decision-making under the coupling of multiple factors [23]. However, these models are encumbered by a substantial number of parameters, and the selection and optimization of these parameters present considerable challenges, thereby hindering their adaptation to multiple dynamic traffic.

**Game theory-based approaches.** Generally, these approaches serve as a popular framework for modeling vehicle interactive behavior, deriving decisions from equilibrium solutions in game models [24], [25]. These models typically strive for equilibrium solutions to determine optimal driving strategies. The types include static/dynamic games, games with complete/incomplete information, and cooperative /non-cooperative games [26], [27]. For example, a decision-making method for AVs based on non-cooperative game theory is presented in [26]. Utilizing an approximate tree-search algorithm, this approach successfully navigated merging scenarios in simulations. Similarly, Zhao et al. [28] introduce a parallel game-based vehicle interaction model, aiming to characterize the nuances of interactive driving behaviors within a social preference framework. These game-theoretic approaches offer the dual advantage of game reasoning and probabilistic theory, enabling them to consider multiple variables that affect vehicle interaction behavior, and can reflect the two-way interaction between vehicles. Nevertheless, they necessitate the empirical determination of the form and parameters of the reward or cost functions, which are relatively fixed specific for the target scenario. This rigidity poses a challenge in adapting and self-learning in accordance with dynamic real-world environments.

*C. Contributions*

In view of the limitations, this paper addresses the challenges in parameters optimization of the reward function in game theory-based models, and the poor adaptability under dynamic conditions. The contributions can be summarized as follows: 1) We propose an adaptive behavioral decision-making framework for AVs in dynamic environments. By offline learning vehicle interactive behavioral characteristics and online identification of optimal matching parameters, the framework enables environmentally adaptive safety decision-making. 2) We develop a vehicle interaction behavior model that considers multiple factors based on non-cooperative game theory. It allows vehicles to engage in behavior interactions in a manner consistent with human logic. Furthermore, it accounts for the individual subjective inner driving pursuits and external traffic constraints, thereby capturing the interaction and conflict process among multiple vehicles. 3) We introduce a parameter optimization method for vehicle interaction behavior model based on maximum entropy IRL. This method leverages interactive behavior feature vectors and probabilistic model outputs, allowing AVs to extract key information and obtain the optimal matching parameters to adapt to dynamic environmental conditions.

*D. Paper Organization*

The remainder of this paper is organized as follows: Section II introduces the overall architecture for our proposed adaptive behavioral decision-making method. The vehicle interaction behavior model based on non-cooperative game for AVs is proposed in Section III. In Section IV, the modelling of behavioral decision-making for AVs in dynamic environment is presented. Subsequently, the performance of the proposed method is evaluated and analyzed in Section V. Also, some discussions about the proposed method are outlined in Section VI. Finally, conclusions of this work are drawn in Section VII.

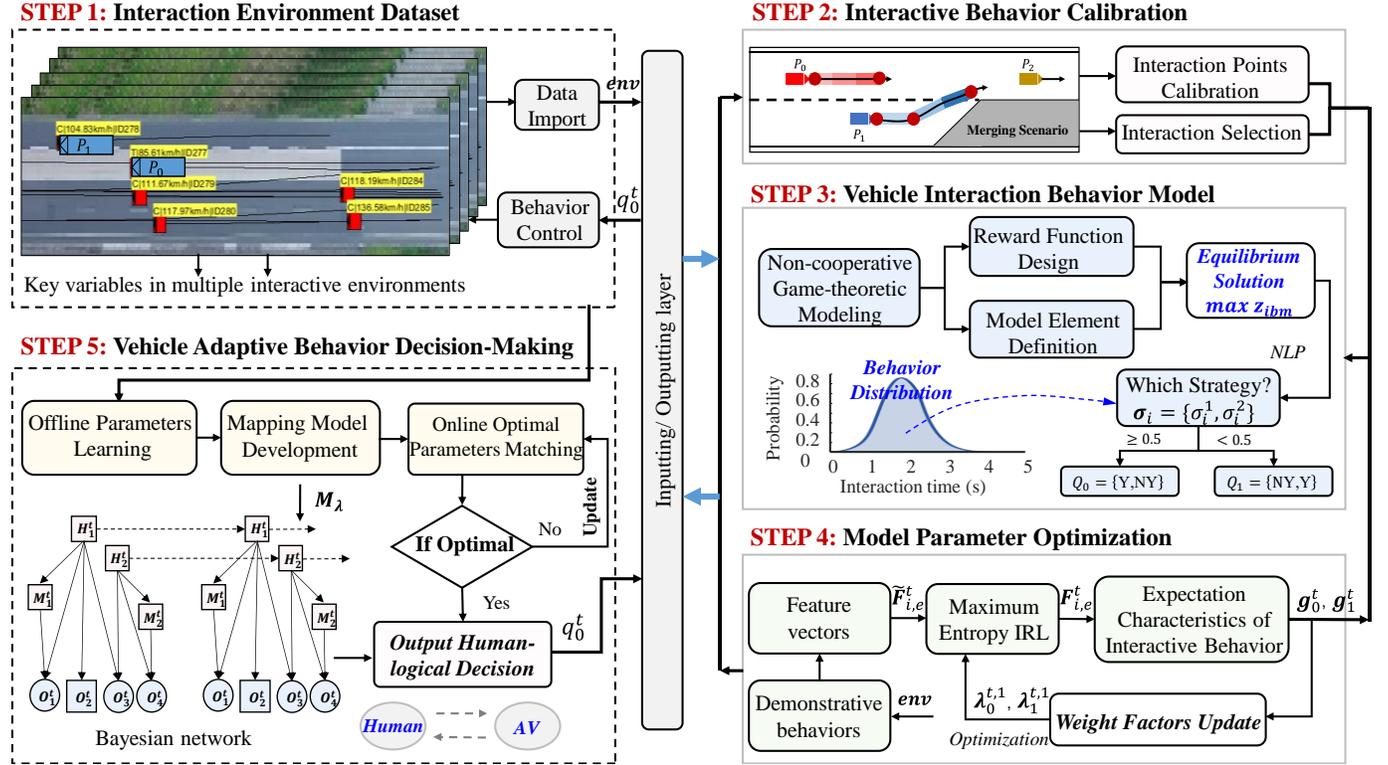

**Fig. 2.** The framework of the proposed adaptive behavioral decision-making for AVs.

## II. FRAMEWORK

**Fig. 2** shows the proposed framework of adaptive behavioral decision-making for AVs in dynamic interactive scenarios. For step 1, key variables like location and speed with vehicle IDs are extracted from the interaction environment dataset. The data is then feed into step 2, during the interactive behavior calibration, data is collected at interaction points based on ramp vehicle lane changes to calibrate their interactive behaviors. A vehicle interaction behavior model is then developed in step 3 employing non-cooperative game theory, identifying influential elements, designing its reward function, and solving for equilibrium in the developed model. Key parameters of this model are optimized in step 4 using Maximum Entropy IRL. This involves establishing feature vectors for vehicle behavior, obtaining behavior probability distributions by maximizing the distribution entropy, and updating weight parameters for model optimization in step 3.

Finally, in step 5, an adaptive behavior decision-making method for AVs in dynamic environments is proposed. This method uses offline learning to determine optimal parameters for various environments, creating a mapping model for online parameter-environment matching. Consequently, in dynamic conditions, the AVs interactive behavior probabilities are generated based on the pre-established vehicle interaction behavior model, enabling crucial model parameters to adaptively adjust in accordance with evolving environmental dynamics. The interactive scenarios with high uncertainty will be established to evaluate the proposed method. Which can demonstrate that the model enables behavioral interaction decisions in a manner consistent with human logic.

## III. VEHICLE INTERACTION BEHAVIOR MODEL

### A. Basic Model Element

In real-world driving scenarios, AVs engage in decision-making due to interactive conflicts with other surrounding vehicles. Game theory effectively models these bidirectional interactions among vehicles, considering diverse various factors within the environment that influence these vehicular interactions. Hence, this study utilizes non-cooperative game theory for modeling such vehicular interactions. This approach is particularly suited for modeling non-cooperative behaviors in scenarios with intense interactions. Which necessitates the application of non-cooperative game-theoretic interaction behavior modeling. As shown in **Fig. 3**, in a dynamic interaction environment, the basic elements in the vehicle interaction behavior model include participants $P$, action space $Q$, reward function $U$, and the hybrid strategy solution $\sigma$. Specifically, $P = \{P_0, P_1\}$, $P_0$ is the ego vehicle and $P_1$ is the interactive vehicle. Given predetermined initiation and termination times for interaction, the main interaction action of the vehicle in the straight interaction phase is longitudinal non-avoidance or avoidance behavior. Consequently, the action space $Q$ is defined as either yielding ("Yield") or non-yielding ("NYield"), articulated as $Q = \{Q_0, Q_1\}$.

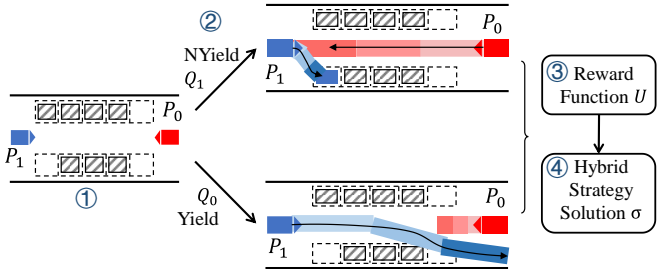

**Fig. 3.** Illustration of basic model elements of vehicle dynamic interaction.

### B. Reward Function Design

Based on real-time traffic information, including vehicle physical states and traffic regulations, a reward function is devised to optimize both driving safety and efficiency. Concurrently, traffic rules and topographical constraints are incorporated to ensure the rationality of interaction outcomes. **Fig. 4** provides a comprehensive analysis of vehicles $P_0$ and $P_1$ during the straight-driving interaction phase, revealing a moderate conflict between $P_0$ and the adjoining ramp vehicle $P_1$ before its apparent lane change to the left.

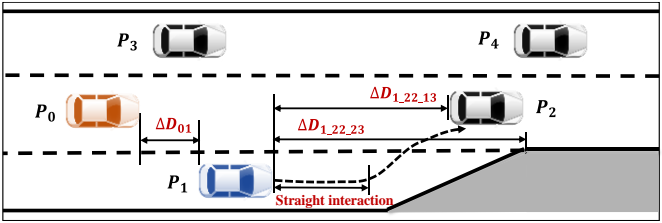

**Fig. 4.** Illustration of vehicle dynamic interaction and model reward function design process.

For vehicle $P_0$, the primary objective is to ensure driving safety. Given the potential interaction conflicts between $P_0$ and $P_1$, and given that vehicles on the main road generally drive at higher speeds, it is crucial to maintain $P_0$'s driving efficiency. Consequently, the reward function $U_0$ for $P_0$ is formulated as follows:

$$U_0 = \lambda_0^1 \frac{\frac{\Delta D_{01}}{v_0}}{t_{\text{norm}}} + \lambda_0^2 \frac{v_0^{\text{des}}}{v_{\text{norm}}} \quad (1)$$

where $\Delta D_{01}$ represents the predicted distance between vehicles $P_0$ and $P_1$ within a forthcoming time interval $\Delta t$. $v_0$ and $v_0^{\text{des}}$ denote the current and feasible speeds within $\Delta t$ of $P_0$ and $P_1$, respectively. The variables $t_{\text{norm}}$ and $v_{\text{norm}}$ serve as the normalization factors for time and speed, respectively. Furthermore, $\lambda_0^1$ and $\lambda_0^2$ function as the weight factors for the reward function $U_0$.

For the ramp vehicle $P_1$, characterized by lane-changing and speed constraints on the ramp, the primary focus is on ensuring its driving safety. The safety of $P_1$ largely depends on the vehicles ahead and the ramp's endpoint, while its rear safety mainly considers the distance between $P_1$ and $P_0$. Accordingly, the reward function $U_1$ for $P_1$ is defined as follows:

$$U_1 = \lambda_1^1 \frac{\frac{\Delta D_{\text{ahead}}}{v_1}}{t_{\text{norm}}} + \lambda_1^2 \frac{\Delta D_{01}}{D_{\text{norm}}} \quad (2)$$

where $\Delta D_{\text{ahead}}$ represents the minimum gap between vehicle $P_1$ and either the vehicles ahead on the main road or the ramp's endpoint, defined as $\Delta D_{\text{ahead}} = \min\{\Delta D_{1\_22\_13}, \Delta D_{1\_22\_23}\}$. If no vehicles are present ahead on the main road, $\Delta D_{\text{ahead}} = \Delta D_{1\_22\_23}$. $v_1$ indicates the current speed of $P_1$, while $D_{\text{norm}}$ serves as the distance normalization factor. Moreover, $\lambda_1^1$ and $\lambda_1^2$ function as the weight factors (main parameters) for the reward function $U_1$.

Specifically, the equation for calculating the predicted distance $\Delta D_{01}$ between vehicles $P_0$ and $P_1$ within the short future time domain $\Delta t$ is as follows:

$$\Delta D_{01} = |\Delta D_{01}^{\text{init}}| + 0.5(|v_1^{\text{des}}| - |v_0^{\text{des}}|) \times \Delta t + 0.5(a_1^{\text{des}} - a_0^{\text{des}}) \times \Delta t^2 \quad (3)$$

where $\Delta D_{01}^{\text{init}}$ denotes the current distance between vehicles $P_0$ and $P_1$. $v_1^{\text{des}}$ and $a_1^{\text{des}}$ are the feasible speed and acceleration within $\Delta t$ of $P_1$, and its calculation Eq. is as follows:

$$v_1^{\text{des}} = v_1 + a_1^{\text{des}} \times \Delta t \quad (4)$$

In particular, the feasible acceleration $a_1^{\text{des}}$ can be calculated according to the behavior specified by $Q_1$, which can be either "Yield" or "NYield." And $\dot{a}_1$ is the jerk for vehicle $P_1$.

$$a_1^{\text{des}} = \begin{cases} a_1 + |\dot{a}_1| \times \Delta t, & Q_1 = \text{NYield} \\ a_1 - |\dot{a}_1| \times \Delta t, & Q_1 = \text{Yield} \end{cases} \quad (5)$$

### C. Solution Methodology for the Interaction Behavior Model

In addressing the solution for the vehicle interaction model, the problem can be reformulated as a nonlinear programming problem. Let the action space of interacting vehicle $P_i$ be $Q_i = \{q_i^1, q_i^2\}$. The mixed strategy for $P_i$ is denoted by $\sigma_i = \{\sigma_i^1, \sigma_i^2\}$, where $\{\sigma_i^k\}_{k=1,2}$ represents the probability of $P_i$ adopting the $k$-th element in its action space $Q_i$. Define $U_i = \left(u_i(\sigma_0^i, \sigma_1^i)\right)_{2\times 2}$ as the payoff matrix for $P_i$, where $u_i(\sigma_0^i, \sigma_1^i)$ is the reward for $P_i$ under the strategy combination $u_i(\sigma_0^i, \sigma_1^i)$. Hence, finding the equilibrium for the game between interacting vehicles $P_0$ and $P_1$ can be translated into the following mathematical programming problem [29]:

$$\begin{aligned}
\max z_{ibm} &= \sigma_0 U_0 \sigma_1^T - v_0 + \sigma_0 U_1 \sigma_1^T - v_1 \\
s.t.\ & U_0 \sigma_1^T \leq v_0 E_2^T \\
& (\sigma_0 U_1)^T \leq v_1 E_2^T \\
& \sigma_0 E_2^T = \sigma_1 E_2^T = 1 \\
& \sigma_0^k \geq 0, k = 1,2 \\
& \sigma_1^k \geq 0, k = 1,2
\end{aligned} \quad (6)$$

where $v_i$ denotes the expected reward for vehicle $P_i$ under the equilibrium strategy, and $E_2^T = [1,1]^T$.

Based on the aforementioned mathematical programming problem, the mixed strategy solutions for the established vehicle interaction model can be obtained, namely $\sigma_0 = \{\sigma_0^1, \sigma_0^2\}$ and $\sigma_1 = \{\sigma_1^1, \sigma_1^2\}$.

In Eq. (6), the objective function for the equilibrium solution in the vehicle interaction behavior model is given by $z_{ibm} = \sigma_0 U_0 \sigma_1^T - v_0 + \sigma_0 U_1 \sigma_1^T - v_1$. Here, the mixed strategies for vehicles $P_0$ and $P_1$ are $\sigma_0 = [\sigma_0^1, \sigma_0^2]$ and $\sigma_1 = [\sigma_1^1, \sigma_1^2]$, respectively. Their corresponding payoff matrices are $\boldsymbol{U}_0 = \begin{bmatrix} u_0^{11} & u_0^{12} \\ u_0^{21} & u_0^{22} \end{bmatrix}, \boldsymbol{U}_1 = \begin{bmatrix} u_1^{11} & u_1^{12} \\ u_1^{21} & u_1^{22} \end{bmatrix}$. Consequently, the objective function can be further derived as follows:

$$z_{ibm} = \sum_{i=1,2, j=1,2} (\sigma_0^i u_0^{ij} \sigma_1^j - v_0) + \sum_{i=1,2, j=1,2} (\sigma_0^i u_1^{ij} \sigma_1^j - v_1) \quad (7)$$

## IV. ADAPTIVE BEHAVIORAL DECISION-MAKING METHOD

Based on the established vehicle interaction behavior model, Section IV further explores the vehicle behavior decision-making method in dynamic traffic environments. The adaptive behavior decision-making framework for AVs is outlined as follows: Initially, optimal weight factors for specific environments are determined by leveraging the interaction behavior model through feature matching for expected interactions. Subsequently, optimal matching parameters are obtained for various time points throughout the interaction phase. Next, environmental variables affecting interaction behavior are identified, and a mapping model between these variables and the model parameters is constructed. Finally, a hybrid approach of offline and online learning realizes adaptive decision-making.

### A. Interactive Behavior Model Weight Factor

According to the reward functions for vehicles $P_0$ and $P_1$ described in Equations (1) and (2), the interaction behavior feature matrices $\boldsymbol{F}_0$ and $\boldsymbol{F}_1$ for both vehicles are characterized as follows:

$$\boldsymbol{F}_0 = \begin{bmatrix} \boldsymbol{f}_0^{11} & \boldsymbol{f}_0^{12} \\ \boldsymbol{f}_0^{21} & \boldsymbol{f}_0^{22} \end{bmatrix}, \boldsymbol{F}_1 = \begin{bmatrix} \boldsymbol{f}_1^{11} & \boldsymbol{f}_1^{12} \\ \boldsymbol{f}_1^{21} & \boldsymbol{f}_1^{22} \end{bmatrix}$$
$$\boldsymbol{\rho}_0 = \left[ \frac{\frac{\Delta D_{01}}{v_0}}{t_{norm}}, \frac{v_{0des}}{v_{norm}} \right], \boldsymbol{\rho}_1 = \left[ \frac{\frac{\Delta D_{ahead}}{v_1}}{t_{norm}}, \frac{\Delta D_{01}}{D_{norm}} \right] \quad (8)$$

where, each component in $\boldsymbol{F}_0$ and $\boldsymbol{F}_1$ is a vector, with $\boldsymbol{\rho}_0$ and $\boldsymbol{\rho}_1$ serving as the vectors used for computing interaction features in the reward functions $U_0$ and $U_1$, respectively. For instance, $\boldsymbol{f}_0^{11}$ denotes the interaction feature $\boldsymbol{f}_0^{11} = [f_0^{11,1}, f_0^{11,2}]$, for vehicle $P_0$ when both $P_0$ and $P_1$ engage in interactions $\{q_0^1, q_1^1\}$. Calculations are made according to $\boldsymbol{\rho}_0$ and variables in Eq.s (3) to (5). Similarly, $\boldsymbol{f}_1^{11}$ represents the analogous feature for $P_1$ under the same interaction conditions, computed based on $\boldsymbol{\rho}_1$ and the key variables, resulting in $\boldsymbol{f}_1^{11} = [f_1^{11,1}, f_1^{11,2}]$.

For example, in a merging scenario, $\boldsymbol{f}_0^{11}$ symbolizes the feature output for $P_0$ when it aims for both safety and efficiency, as $P_0$ and $P_1$ engage in interactions defined by $\{\boldsymbol{\sigma}_0^1, \boldsymbol{\sigma}_1^1\}$, represented as $\boldsymbol{\sigma}_0$ and $\boldsymbol{\sigma}_1$. In behavior interaction model, the weight factors for $P_0$ and $P_1$ are designated as $\boldsymbol{\lambda}_0 = [\lambda_0^1, \lambda_0^2]$ and $\boldsymbol{\lambda}_1 = [\lambda_1^1, \lambda_1^2]$, respectively. Each factor is associated with a corresponding interaction feature vector. The matrices of interaction features linked to $\boldsymbol{\lambda}_0 = [\lambda_0^1, \lambda_0^2]$ and $\boldsymbol{\lambda}_1 = [\lambda_1^1, \lambda_1^2]$ are illustrated as follows:

$$\begin{aligned} \lambda_0^1 &\rightarrow \begin{bmatrix} f_0^{11,1} & f_0^{12,1} \\ f_0^{21,1} & f_0^{22,1} \end{bmatrix}, \lambda_0^2 \rightarrow \begin{bmatrix} f_0^{11,2} & f_0^{12,2} \\ f_0^{21,2} & f_0^{22,2} \end{bmatrix} \\ \lambda_1^1 &\rightarrow \begin{bmatrix} f_1^{11,1} & f_1^{12,1} \\ f_1^{21,1} & f_1^{22,1} \end{bmatrix}, \lambda_1^2 \rightarrow \begin{bmatrix} f_1^{11,2} & f_1^{12,2} \\ f_1^{21,2} & f_1^{22,2} \end{bmatrix} \end{aligned} \quad (9)$$

Consequently, the objective function $z_{ibm}$ in the vehicle interaction model can be construed as a function composed of weight factors $[\boldsymbol{\lambda}_0, \boldsymbol{\lambda}_1]$ and interaction features $[\boldsymbol{F}_0, \boldsymbol{F}_1]$. Given that all interacting vehicles aim to maximize their respective reward functions; the probability distribution of interactions can be modeled as an exponential probability distribution. Within this framework, each element in the feature vector will exhibit a linear relationship with its corresponding interaction behavior.

### B. Model Parameter Optimization via Maximum Entropy IRL

In dynamic interactive traffic scenarios, to achieve optimal matching parameters between the interaction behavior model and the environment, optimizing weight coefficients in its reward functions is crucial. Specifically, given a set of demonstrative interactions under current conditions, we aim to obtain a probability distribution $P(\sigma_h)$ such that the expected features of interactions under this distribution closely align with, or consistent with, the empirical expected features ($\tilde{f}_h$), denoted as $\mathrm{E}_{P(\sigma_h)}[f_h] = \tilde{f}_h$. Currently, reward function design often relies on empirical methods. However, Maximum Entropy IRL based on behavior feature matching can derive crucial parameters in the reward function based on expert or demonstrative actions [30], [31]. This helps in understanding the rules or mutual influences between generating certain actions or policies and reward function parameters. Hence, this paper proposes a parameter optimization method based on Maximum Entropy IRL. Utilizing this optimization approach with expert demonstrations refines the model parameters, specifically the weight factors, ensuring the model's behavioral output closely imitates demonstrative actions.

To optimize vehicle behavior, we first obtain the environmental information $\boldsymbol{E}_0^t$ and $\boldsymbol{E}_1^t$ for vehicles $P_0$ and $P_1$ at a specific time $t$. This includes key metrics like relative distance and speed. Concurrently, we capture the demonstrative behaviors $\tilde{q}_0^t$ and $\tilde{q}_1^t$ from the human-driven vehicles in the dataset and compute their feature vectors $\widetilde{\boldsymbol{F}}_0^t$ and $\widetilde{\boldsymbol{F}}_1^t$.

Next, we initialize the weight factors $\boldsymbol{\lambda}_0^{t,0}$ and $\boldsymbol{\lambda}_1^{t,0}$ in the reward functions for $P_0$ and $P_1$. Utilizing the established vehicle interaction model and the initialized reward functions $U_0^t$ and $U_1^t$, the model equilibrium solving module computes the behavior probability distribution or mixed strategies $\boldsymbol{\sigma}_0^t$ and $\boldsymbol{\sigma}_1^t$, along with their corresponding interaction feature vectors $\boldsymbol{F}_0^t$ and $\boldsymbol{F}_1^t$. For instance, each element in $\boldsymbol{F}_0^t$ represents the result calculated by feature metric $\boldsymbol{\rho}_0^t$ when $P_0$ and $P_1$ adopt behaviors from their respective action spaces $\boldsymbol{Q}_0$ and $\boldsymbol{Q}_1$. $\boldsymbol{F}_1^t$ is calculated similarly, using $\boldsymbol{\rho}_1^t$ as indicated in Eq. (10). $\boldsymbol{\rho}_0^{q_0^1, q_1^1}$ is the interaction feature vector obtained from the $\boldsymbol{\rho}_0$ metric when $P_0$ selects action $q_0^1$ and $P_1$ selects action $q_1^1$.

$$\boldsymbol{F}_0^t = \begin{bmatrix} \boldsymbol{u}_0^{q_0^1, q_1^1} & \boldsymbol{u}_0^{q_0^1, q_1^2} \\ \boldsymbol{u}_0^{q_0^2, q_1^1} & \boldsymbol{u}_0^{q_0^2, q_1^2} \end{bmatrix}, \boldsymbol{F}_1^t = \begin{bmatrix} \boldsymbol{u}_1^{q_1^1, q_0^1} & \boldsymbol{u}_0^{q_1^2, q_0^1} \\ \boldsymbol{u}_0^{q_1^1, q_0^2} & \boldsymbol{u}_0^{q_1^2, q_0^2} \end{bmatrix} \quad (10)$$

Next, using the behavior probability distributions $\boldsymbol{\sigma}_i^t$ output by the vehicle interaction model, the expected interaction features $\boldsymbol{F}_{i,e}^t$ for vehicle $P_i$ are calculated. For instance, the expected expression for $\boldsymbol{\rho}_0^{q_0^1,q_1^1}$ can be computed as $\boldsymbol{\sigma}_0^t(q_0^1)\boldsymbol{u}_0^{q_0^1,q_1^1}\boldsymbol{\sigma}_1^t(q_1^1)$. In a similar fashion, the empirical features $\widetilde{\boldsymbol{F}}_{i,e}^t$ for the demonstration behavior are determined.

Lastly, based on the expected interaction features and the expected features of the demonstration behavior, the feature gradient vectors for vehicles $P_0$ and $P_1$ can be obtained.

$$\boldsymbol{g}_0^t = \widetilde{\boldsymbol{F}}_{0,e}^t - \boldsymbol{F}_{0,e}^t \\ \boldsymbol{g}_1^t = \widetilde{\boldsymbol{F}}_{1,e}^t - \boldsymbol{F}_{1,e}^t \quad (11)$$

Utilizing the feature gradient vectors $\boldsymbol{g}_0^t$ and $\boldsymbol{g}_1^t$ obtained from Eq. (11), the weight factors of the vehicle interaction model are updated via Eq. (12). This yields new weight factors $\boldsymbol{\lambda}_0^{t,1}$ and $\boldsymbol{\lambda}_1^{t,1}$ for vehicles $P_0$ and $P_1$, respectively. And $\delta$ represents the feature gradient update parameter.

$$\boldsymbol{\lambda}_0^{t,1} = \boldsymbol{\lambda}_0^{t,0} - \delta \times \boldsymbol{g}_0^t \\ \boldsymbol{\lambda}_1^{t,1} = \boldsymbol{\lambda}_1^{t,0} - \delta \times \boldsymbol{g}_1^t \quad (11)$$

The aforementioned steps are iteratively executed until the feature gradient vectors fall below a predetermined threshold ε. At that point, the optimization process terminates, and the refined weight factors are considered as the optimal factors $\boldsymbol{\lambda}_0^t$ and $\boldsymbol{\lambda}_1^t$ for vehicles $P_0$ and $P_1$ in the reward functions of the vehicle interaction model at time $t$. The optimization and iterative process for the weight factors, based on the expected interaction feature matching, are illustrated as **Algorithm 1**.

**Algorithm 1** Model Parameters Optimization Method

**Input:** Demonstrative behaviors $\tilde{q}_0^t$ and $\tilde{q}_1^t$ at $t$; empirical features of the demonstrative behaviors $\widetilde{\boldsymbol{F}}_{0,e}^t$ and $\widetilde{\boldsymbol{F}}_{1,e}^t$; initial weight factors $\boldsymbol{\lambda}_0^{t,0}$ and $\boldsymbol{\lambda}_1^{t,0}$; convergence threshold $\varepsilon$; feature gradient update parameter $\delta$;
**Output:** Optimal weight factors $\boldsymbol{\lambda}_0^t$ and $\boldsymbol{\lambda}_1^t$ for vehicles $P_0$ and $P_1$ in the reward functions;
1: $\boldsymbol{\lambda}_0 \leftarrow \boldsymbol{\lambda}_0^{t,0}; \boldsymbol{\lambda}_1 \leftarrow \boldsymbol{\lambda}_1^{t,0}$;
2: **While** $\|\boldsymbol{g}_0^t\|_2 > \varepsilon$ and $\|\boldsymbol{g}_1^t\|_2 > \varepsilon$ **do**
3:   $[\boldsymbol{\sigma}_0^t, \boldsymbol{\sigma}_1^t] = \arg\max z_{ibm}$;
4:   Calculate the interaction feature vectors $\boldsymbol{F}_0^t$ and $\boldsymbol{F}_1^t$ for $P_0$ and $P_1$ based on Eq. (10);
5:   Calculate the expected interaction features $\boldsymbol{F}_{0,e}^t$ and $\boldsymbol{F}_{1,e}^t$ for $P_0$ and $P_1$ according to $[\boldsymbol{\sigma}_0^t, \boldsymbol{\sigma}_1^t]$;
6:   Calculate the feature gradients of $\boldsymbol{\lambda}_0$ and $\boldsymbol{\lambda}_1$ based on Eq. (11): $\boldsymbol{g}_0^t = \widetilde{\boldsymbol{F}}_{0,e}^t - \boldsymbol{F}_{0,e}^t$; $\boldsymbol{g}_1^t = \widetilde{\boldsymbol{F}}_{1,e}^t - \boldsymbol{F}_{1,e}^t$;
7:   $\boldsymbol{\lambda}_0 \leftarrow \boldsymbol{\lambda}_0 - \delta \times \boldsymbol{g}_0^t$;
8:   $\boldsymbol{\lambda}_1 \leftarrow \boldsymbol{\lambda}_1 - \delta \times \boldsymbol{g}_1^t$;
9: **Return** $\boldsymbol{\lambda}_0^t, \boldsymbol{\lambda}_1^t$

*C. Vehicle Adaptive Behavior Decision-Making Process*

Considering the challenges posed by the dynamic adaptability of vehicle interaction behaviors, we propose a hybrid approach that combines offline and online learning for adaptive vehicle behavior decision-making, as illustrated in **Fig. 5**. Within the dynamic interactive traffic environment, based on the dynamic traffic observation variables $\{O_1, O_2, O_3\}$, the traffic participants $\{P_0, P_1\}$, and the demonstrative behaviors $\tilde{q}_0^t$, $\tilde{q}_1^t$, then the interaction behavior probability $\{\sigma_0^t, \sigma_1^t\}$ is computed through a combination of offline and online learning phases, ultimately leading to the output of vehicle adaptive decision-making behaviors [32]. Specifically, observable variables $\{O_1, O_2, O_3\}$ represent dynamic environmental information; based on the extraction of key environmental information, observation node $O_1$ represents $\Delta d_{01}^y$ and $\Delta v_{01}^x$, $O_2$ represents $\Delta d_{01}^x$, and $O_3$ represents $\Delta D_{\text{ahead}}$ and $\Delta v_1^x$.

During the offline learning phase, we initially employ weight factor learning for expected feature matching of vehicle interactive behaviors. Subsequently, based on the calibrated results, we utilize a Maximum Entropy IRL algorithm to optimize model parameters, thereby achieving an optimal parameterization for modeling vehicle interactions. The output is adjusted with the real-time environmental information $E^t$, and then feed into online learning phase.

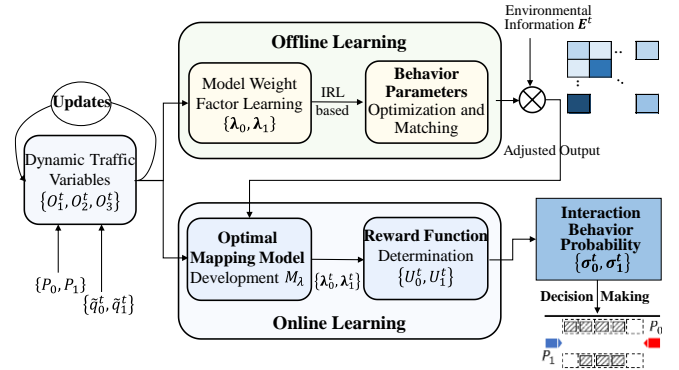

**Fig. 5.** Vehicle adaptive behavior decision-making process based on offline and online learning.

During the online learning phase, to align the dynamic driving environment with the optimal weight factors of the vehicle interaction behavior model and output the interaction behavior probabilities for decision-making, we initially establish a mapping model $M_\lambda$ from environmental variables to the optimal matching weight factors. Given the inherent uncertainty in environmental variables, we consider employing a Bayesian network to fit the uncertain data [33], [34], thereby establishing a mapping model network structure as illustrated in **Fig. 6**. In this model, latent variables $H$ and $H_2$ represent the weight factors $\boldsymbol{\lambda}_0$ and $\boldsymbol{\lambda}_1$ of vehicles $P_0$ and $P_1$, respectively. Latent variables $M_1$ and $M_2$ denote mixed Gaussian parameters. Observable variables $\{O_1, O_2, O_3\}$ is dynamic environmental information. Utilizing training samples that furnish environmental variables and corresponding weight factors under multiple distinct interaction environments, optimal parameterization for $M_\lambda$ is achieved through normalization and discretization of weight factors. Upon obtaining the parameters in the mapping model, probabilistic inference yields the weight factors' distribution $\{\boldsymbol{\lambda}_0^t, \boldsymbol{\lambda}_1^t\}$ for the given environmental variables $\{O_1^t, O_2^t, O_3^t\}$.

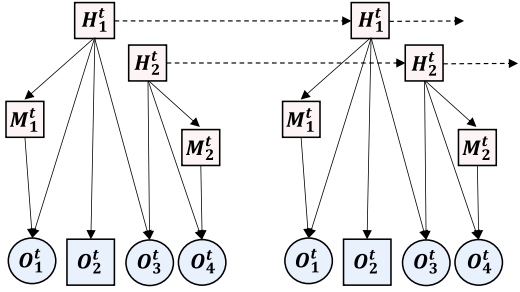

**Fig. 6.** The structure of mapping model based on Bayesian network.

Leveraging the mapping model $M_\lambda$, $P_0$ performs real-time recognition of the current traffic at time $t$. It extracts the critical environmental data $\{\Delta d_{01}^x, \Delta d_{01}^y, \Delta v_{01}^x, \Delta D_{\text{ahead}}, \Delta v_1^x\}$ required for observed variables $\{O_1, O_2, O_3\}$. This yields optimal matching weight factors $\{\lambda_0^t, \lambda_1^t\}$ for the reward function $U_0^t$ and $U_1^t$. Solving the vehicle interaction model results in the behavior probabilities $\sigma_0^t$ for $P_0$ and $\sigma_1^t$ for $P_1$ at the subsequent time step. Consequently, $P_0$ can select its next interaction behavior, $q_0^t$, based on the maximum likelihood in its mixed strategy $\boldsymbol{\sigma}_0^t$.

## V. Experiments and Results

In this section, to verify and test the performance of proposed adaptive behavioral decision-making method, we first focus on data preparation and preprocessing. Subsequently, publicly available datasets are employed to validate the feasibility and effectiveness of the proposed model. Furthermore, we tested the performance based on the data collected from the real vehicle platform in typical interactive scenarios.

### A. Data Preparation and Preprocessing

We first focus on data preparation and preprocessing. The data validation employs large-scale naturalistic vehicle trajectory datasets collected from multiple segments of German highways, including the highD and exiD datasets [35], [36]. These datasets record lane-changing or ramp-merging scenarios at a frame rate of 25 Hz (every 40 ms). Additionally, the datasets utilize quadcopters equipped with high-definition cameras, as illustrated in **Fig. 7**, to capture the motion data of each vehicle from a bird's-eye perspective. The datasets can measure each vehicle's position and movement states on the road to acquire both lateral and longitudinal positional information, with a localization error less than 10 cm.

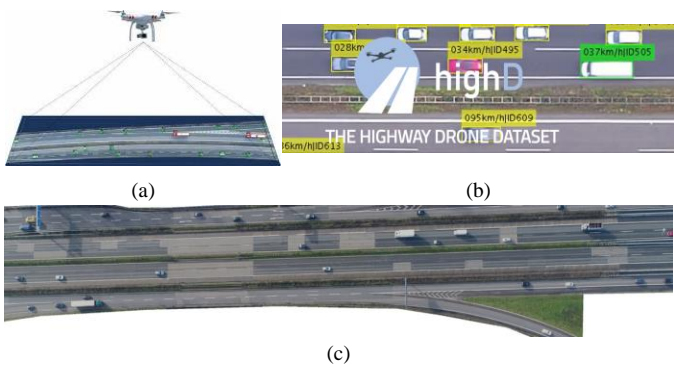

**Fig. 7.** Dataset Preparation. (a) dataset collection, (b) highD dataset and (c) exiD dataset.

Based on these datasets, vehicle data is extracted under typical ramp-merging scenarios. Initially, ramp vehicle IDs are obtained from their lane IDs. Subsequently, the key lane-change moments for these vehicles are identified to isolate interacting main carriageway vehicles and their data information is extracted. Following data preprocessing, we proceed to validate a parameter optimization method, predicated on interactive behavior expectation feature matching.

In the preprocessing phase, interaction start and end moments in ramp-merging scenarios are defined. The interaction endpoint is calibrated using evolving lane-changing probabilities. For example, ramp vehicle ID=782, shown in **Fig. 8**, has its driving intent analyzed through a pre-established model [33], [34]. **Fig. 9** presents the calibrated results and associated recognition probabilities. Specifically, the moment when the ramp vehicle's lane-changing probability exceeds 50%, without being overtaken by main carriageway vehicles, marks the interaction endpoint.

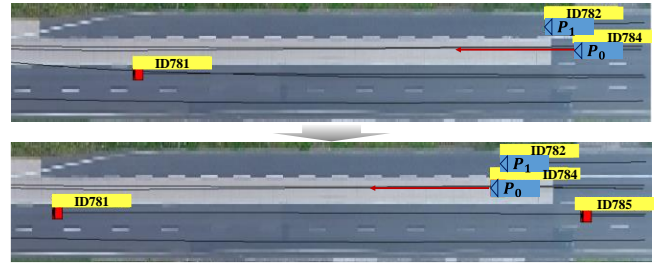

**Fig. 8.** The state evolution of the traffic environment with vehicle ID 782.

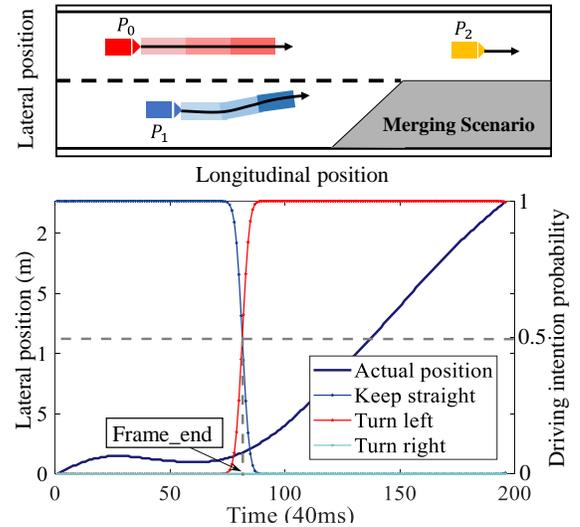

**Fig. 9.** Interaction time calibration of vehicle ID 782.

Finally, the interactive behaviors between the main carriageway vehicle $P_0$ and the ramp vehicle $P_1$ are calibrated. The calibration criterion for interactive behavior is defined as follows: based on the longitudinal acceleration value $a_x$, if the acceleration is negative, the behavior is calibrated as yielding; if positive, further calibration is conducted based on its jerk $\dot{a}_x$. If the jerk is positive, the behavior is calibrated as non-yielding; otherwise, it is calibrated as yielding. The calibrated results for the interactive behavior between $P_0$ and $P_1$ are illustrated in **Fig.**

**10**. **Fig. 10**(a) presents the acceleration and calibration outcomes, which primarily rely on the jerk variations shown in **Fig. 10**(b). The red and purple curves in **Fig. 10**(a) represent the non-yielding probabilities for $P_1$ and $P_0$, respectively.

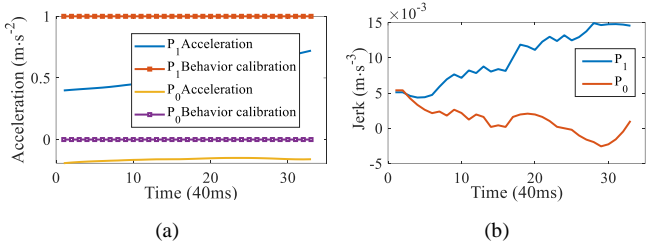

**Fig. 10.** Interaction behavior calibration of vehicle ID 782. (a) Acceleration and calibration results and (b) jerk description.

Utilizing the highD and exiD datasets, post-preprocessing yields a total of 1738 interactive vehicle data sequences. These sequences encapsulate environmental variables throughout each interaction, as well as the calibrated interactive behaviors of the vehicles involved.

### B. Verification of Model Parameter Optimization Method

Based on the processed environmental variables and the demonstrated interactive behavior samples, we extract driving environmental information at each instance for every interactive data sequence. Leveraging the parameter optimization method predicated on interactive behavior expectation feature matching, and employing the calibrated interactive behaviors of vehicle $P_0$ and ramp vehicle $P_1$, optimal matching parameters for the vehicle interaction model are determined at each time point. This method comprehensive parameter optimization within each interactive sequence.

Using vehicle $P_1$ with ID 782 in **Fig. 8** as an example, the optimized matching parameters are inputted into the model's reward function. As illustrated in **Fig. 11**, the optimized parameters yield outputs consistent with the demonstrated behaviors, thereby technically reconstructing the demonstrated actions (as compared to the calibrated behaviors shown in **Fig. 10**(a)). This confirms the model's capability to facilitate AVs in learning human decision-making in interactive scenarios. It also highlights the pivotal role of reward function parameters, indicating that the proposed optimization technique allows for adaptive learning in various driving environments.

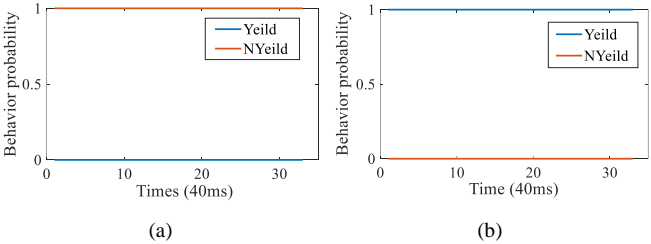

**Fig. 11.** Interaction model parameter learning results for vehicle 782. (a) Learning results of vehicle $P_1$ and (b) learning results of vehicle $P_0$.

### C. Verification of Adaptive Behavior Decision-Making Methods in Dynamic Environments

The obtained model parameters serve dual functions: 1) to corroborate the efficiency of the optimization method predicated on interactive behavior feature matching, and 2) to yield data samples of optimal matching parameters under diverse environmental conditions. Therefore, these samples not only provide training data for subsequent models but also enable validation of interactive decision-making methods.

#### 1) Results and Comparative Analysis

The parameter optimization method involves the post-processing of 1738 sequence data in environmental variables and demonstration interaction behavior data samples. Each sequence contains corresponding environmental variables, demonstration interaction behaviors, and optimally matched model parameters throughout the interaction process. A subset of 1,550 samples is allocated for training the mapping model, and 188 for testing. The training set comprises 65,198 data points, each corresponding to distinct environmental variables and optimal parameters. These are used to train the mapping model, which is subsequently employed to validate vehicle behavior decision-making methods in dynamic traffic.

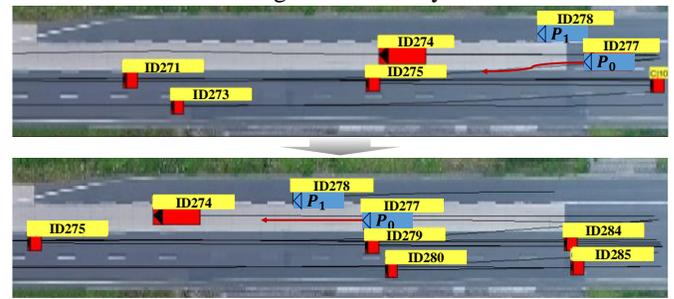

**Fig. 12.** The state evolution of the traffic environment with vehicle ID 278 (blue) approaching the interactive vehicle in the merging zone.

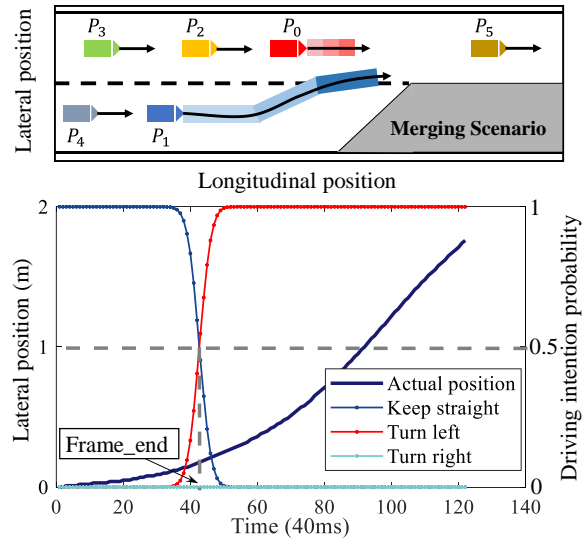

**Fig. 13.** Interaction time calibration of vehicle ID 278.

**Fig. 12** reveals the evolving driving conditions for the main carriageway vehicle $P_0$ and ramp vehicle $P_1$ (ID 278) during $P_1$' straight-driving phase. Upon phase completion, $P_1$ executed a lane change ahead of $P_0$. Also, according to the established criteria for interactive moments and calibrated interactive behaviors, the case of vehicle $P_1$ (ID 278) within this specific interaction scenario serves as an illustrative example. The calibrated outcomes for its interactive termination moment and interactive behaviors are respectively exhibited in **Fig. 13** and **Fig. 14**. During the interaction, environmental variables at distinct instances feed into the mapping model, facilitating the extraction of adaptive parameters for the vehicle interaction behavior model. This informs the model's reward function, generating behavior probabilities for both the ego and interacting vehicles, as showcased in **Fig. 15**.

Compared with the interactive behavior calibration results in **Fig. 14**, the proposed vehicle adaptive behavior decision-making method, which integrates offline and online learning, is capable of learning human-driven vehicle interactions in dynamically changing environments. It generates behavior decision outcomes that exhibit high similarity to those of human-driven vehicles in interactive scenarios.

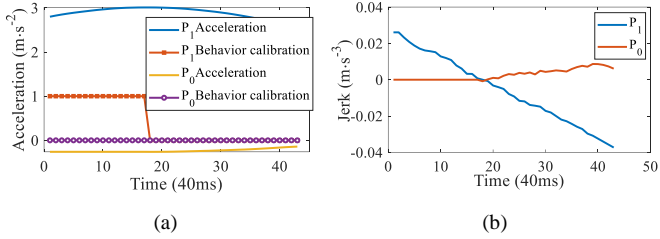

**Fig. 14.** Interaction behavior calibration of vehicle ID 278. (a) Acceleration and calibration results and (b) jerk description.

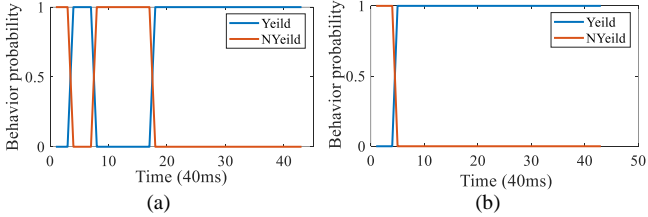

**Fig. 15.** Interaction model parameter learning results for vehicle ID 278. (a) Learning results of vehicle $P_1$ and (b) learning results of vehicle $P_0$.

*2) Performance Evaluation of Validation Results*

To further substantiate the efficiency of the proposed adaptive behavior decision-making method, we employ vehicle behavior interaction model behavior decision outcomes under fixed parameters as comparison benchmarks. As shown in **Fig. 16**, these outcomes are obtained under fixed parameters "[0.8,0.2,0.8,0.2]" and "[0.2,0.8,0.2,0.8]" respectively. The results reveal that the proposed method achieves the highest similarity to calibrated interactive behavior outcomes, thus effectively emulating human-like driving in interactions.

Finally, employing a similar validation procedure and focusing on a vehicle with ID 429 in the interaction scenario, we calibrated its interactive behavior based on its acceleration and jerk metrics, shown in **Fig. 17**. Leveraging the mapping model, we determined optimal model parameters under diverse environmental conditions. The resulting behavior decisions of ego vehicle and predicted interactions are presented in **Fig. 18**. When compared to the calibrated behaviors in **Fig. 17**(a), it becomes evident that our dynamically adaptive vehicle interaction behavior decision method achieves a high degree of similarity to actual human-driven vehicle behavior.

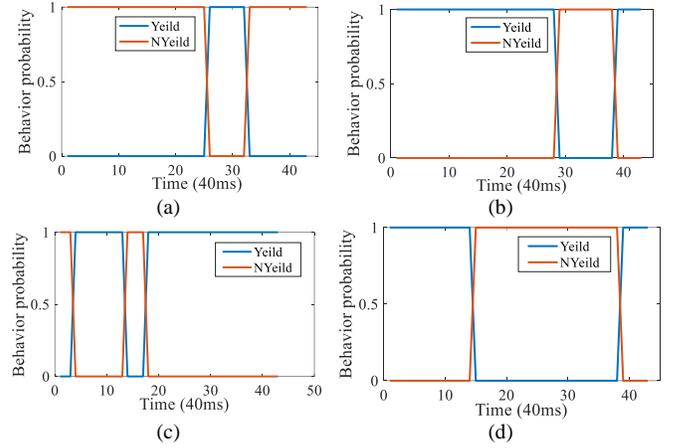

**Fig. 16.** Results of vehicle interaction behavioral model under the first/second fixed parameters. (a) (c) Behavioral decision-making results of $P_1$ and (b) (d) behavior prediction results of $P_0$.

To evaluate the effectiveness of our dynamic, adaptive decision-making method, which employs parameter optimization to learn human driving behavior in dynamic environments, we introduced two metrics:

1) Human-like decision similarity [37], [38]: This measures the similarity between the behavior probabilities output by the vehicle interaction model and the calibrated real human driving behavior. By collecting the number of matching instances across all interactive scenarios and normalizing this by the total number of instances, we obtain a representation of the model's learning and imitation capabilities.

2) Adaptive similarity in varied interaction scenarios [39]: This assesses the model's real-world adaptability by measuring its similarity with calibrated human behaviors in dynamically changing interaction tests.

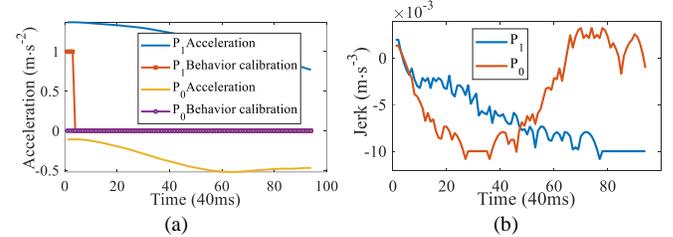

**Fig. 17.** Interaction behavior calibration of vehicle ID 429. (a) Acceleration and calibration results and (b) jerk description.

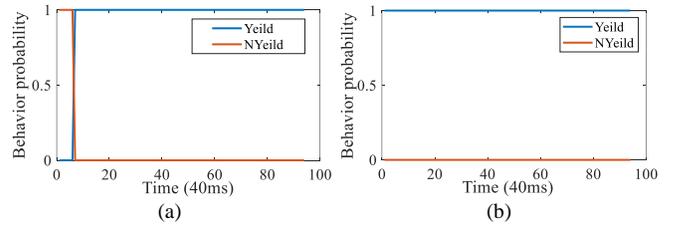

**Fig. 18.** Interaction model parameter learning results for vehicle ID 429. (a) Learning results of vehicle $P_1$ and (b) learning results of vehicle $P_0$.

As shown in **Table 1**, from the highD and exiD datasets, we extracted 11,583 data points across 188 test sequences. Of these, 145 sequences showcased interaction dynamics, accounting for 8,964 data points. Our proposed method achieved an 81.73% human-like decision similarity, corresponding to 9,467 matching points. Specifically, in the highD subset, the similarity rate was 83.12% over 2,316 points, with 1,925 matches. These findings affirm our method's effectiveness in learning and imitating human driving behavior in dynamic environments. Regarding adaptability in fluctuating interaction scenarios, the model exhibited a 77.12% similarity rate with 6,913 matching points. This validates the model's capacity for environmentally adaptive decision-making analogous to human driving behavior.

TABLE I.    MODEL PERFORMANCE IN DIFFERENT SCENARIOS

| Scenario | Sequence | Matching point | Similarity rate |
| --- | --- | --- | --- |
| highD+exiD | 188 | 9467/11583 | 81.73% |
| highD | 38 | 1925/2316 | 83.12% |
| Varied interactive scenarios | 145 | 6913/8964 | 77.12% |

D. *Model Verification based on Real Vehicle Platform*

To further validate the effectiveness and applicability of the proposed model, we established a multi-vehicle experiment platform, employing both real-time and offline validation.

*1) Multi-Vehicle Experimental Platform*

To acquire real-world vehicle data, three test vehicles including the Toyota Prius, Buick GL8, and BMW i3, were deployed, with configurations as described in **Fig. 19**. Specialized hardware, encompassing an R60S-U receiver, dual satellite antennas, D20 networking module, and other hardware devices, output data on geographic coordinates, velocity, orientation, etc. Video data captured through onboard systems provide additional context, such as lane occupancy.

*2) Implementation Process*

To faithfully replicate scenarios present in the dataset, the highway scenarios are selected for validation, showing high congruence with highD scenarios (**Fig. 19**). Furthermore, we assign three test vehicles to respectively assume the roles of Vehicle A, B, and C. Each test vehicle is operated by a driver with a rich driving experience and a distinctive driving style. Additionally, two technical operators are allocated to each vehicle to monitor data quality and manage the equipment. Drivers execute the repeated straight, left, and right lane-change maneuvers. The data obtained for analysis includes the vehicle motion state data outputted by positioning devices and video data recorded by video recording equipment. After statistical analysis, 13 complete and usable real-lane-change datasets were identified. The overall driving speed of the vehicles was around 80 kilometers per hour, meeting the requirements for method validation. Specifically, the validation steps for the vehicle interaction decision model are:

- Preprocessing of motion data for interacting vehicles A and B, and B's lead vehicle;
- Derive relevant operational and environmental variables as model input;
- Select and calibrate vehicles' interaction behaviors;
- Apply the model to generate behavior probabilities at varied time points, subsequently comparing these with calibration metrics to evaluate model's performance.

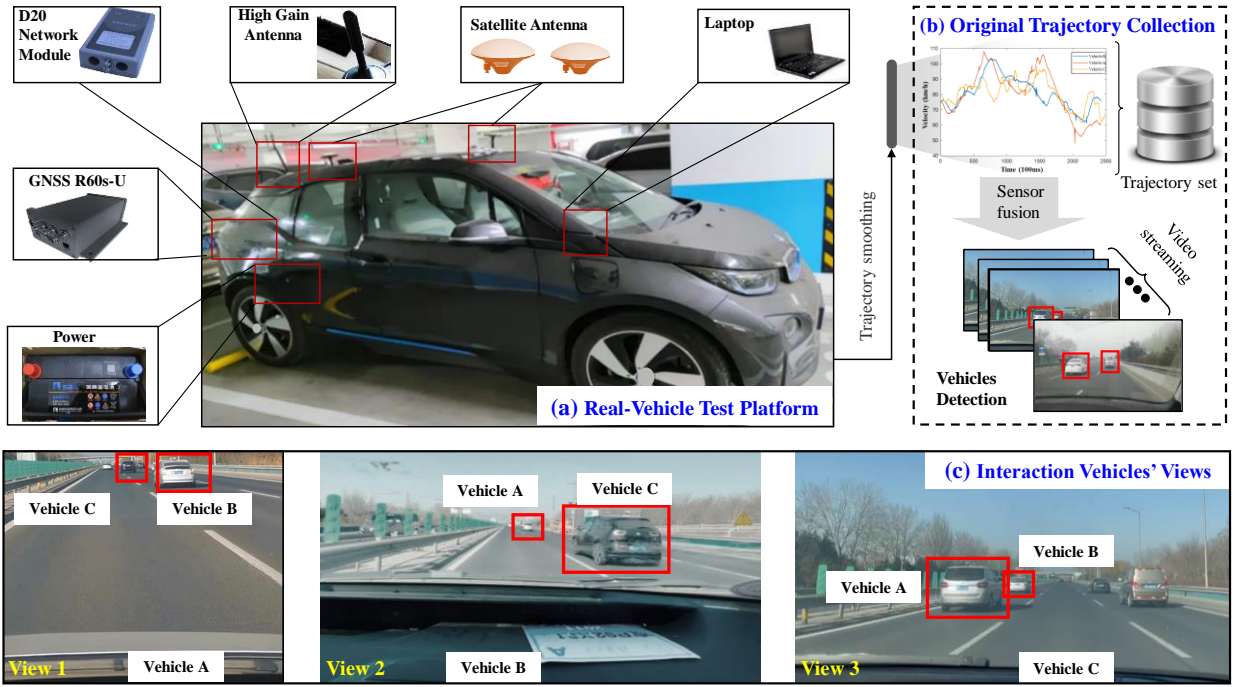

**Fig. 19.**   Real-vehicle test platform and road section overview.

*3) Results Analysis*

**Fig. 20** illustrates the spatiotemporal distribution for the three vehicles under the interactive scenario. The whole verification process can be divided into three stages. Initially, based on the driving intention recognition probabilities, the interaction termination frame can be determined when the lane-changing probability exceeds 50%. As illustrated in **Fig. 21**, the interaction endpoint can be subsequently defined. Then, calibration rules are applied to calibrate the probability of non-yielding interaction behavior between the lane-changing vehicle $P_0$ and the conflict vehicle $P_1$ in the target lane. The calibration outcomes are shown in **Fig. 22**.

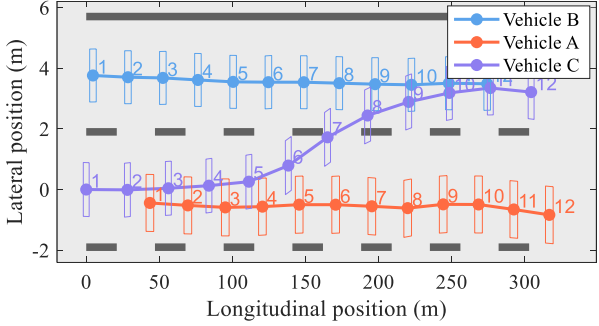

**Fig. 20.** The spatiotemporal distribution for the three vehicles

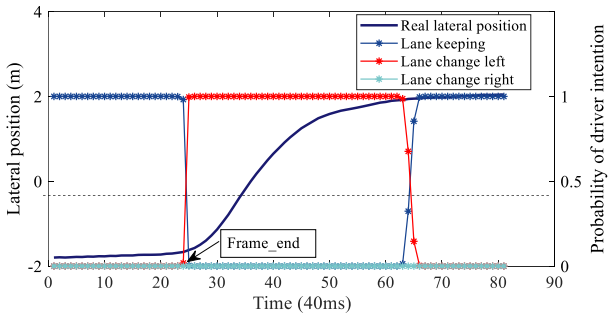

**Fig. 21.** The illustration of determining the interaction termination moment.

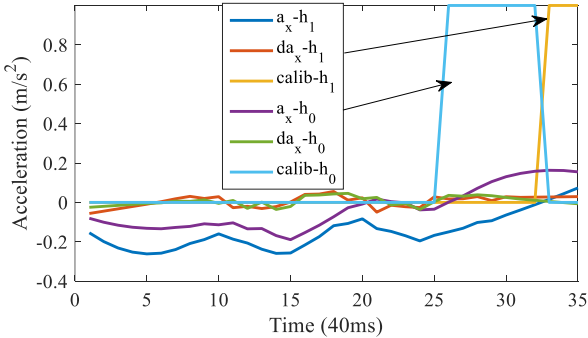

**Fig. 22.** The calibration process of non-yielding behavior probability of interactive vehicle.

Finally, employing the proposed vehicle adaptive decision-making method, the output behavior decisions are presented in **Fig. 23**(a), (b). Additionally, three distinct sets of fixed weights were embedded, one example of the yielding interaction behavior probabilities is as shown in **Fig. 23**(c), (d). It can be seen that the output results of the proposed method are closer and similar to the calibration behavior of human drivers.

In the interactive decision-making process, the proposed method may bring risks of safety violations due to the output decision-making strategies, which is not ideal in real traffic. Safety is a crucial aspect of vehicle interaction, and we then evaluate performance using the 'safety violation rate', serves as an indicator of the vehicle's behaviors' safety level [39]. As shown in **Table 2**, based on the analysis of interactive decision-making results between the lane-changing vehicle and the conflict vehicle in the target lane, the proposed method exhibited a human-like decision-making similarity of 72.73% in the test scenarios and 69.84% in variable interactive scenarios. Simultaneously, our strategy demonstrates exceptional safety, with no safety violations occurring throughout the real-vehicle experimental interaction process. This indicates a potential reduction in the incidence of collisions and severe conflicts.

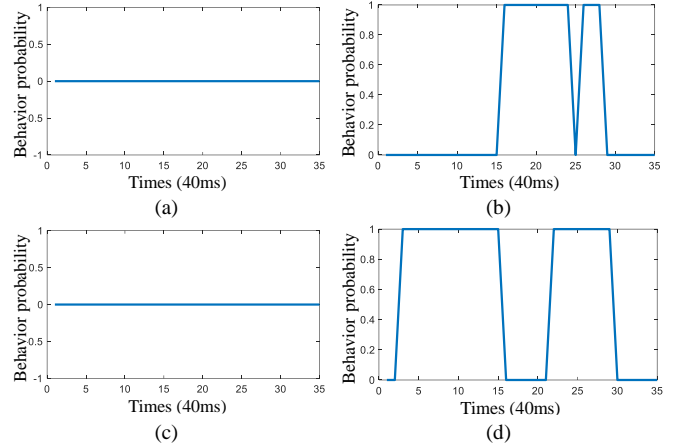

**Fig. 23.** Results of vehicle non-yielding interaction behavior probability under three fixed parameters. (a) (c) Non-yielding behavior probability of lane-changing vehicle $P_0$ and (b) (d) non-yielding behavior probability of conflict vehicle $P_1$.

TABLE II.  MODEL PERFORMANCE IN REAL VEHICLE EXPERIMENTS

| Scenario | Sequence | Similarity rate | Safety violation rate |
|---|---|---|---|
| Test scenarios | 25 | 72.73% | 0% |
| Varied interactive scenarios | 13 | 69.84% | 0% |

In the ramp merging scenarios within the highD dataset, interactions mainly occur due to the obligatory lane-changing by ramp vehicles. Replicating such scenarios in real-world highway tests poses challenges. Specifically, external high-speed vehicles may disrupt test conditions, and inherent coordination among real-vehicle drivers constrains validation. Despite these limitations, comparative validation results demonstrate the applicability of the proposed interactive decision-making method in real-world highway settings.

## VI. DISCUSSION

In this paper, we set out to find an adaptive behavior decision-making method for AVs, applying it to dynamic interaction scenarios. This method was validated and analyzed in real-world vehicle experiments and simulations, assessing its human-like decision-making similarity, safety compliance, and adaptability. This evaluation focused on its effectiveness in interactive behaviors across different traffic environments.

**1) Various interactive scenarios.** Although the verification scenarios in this paper primarily focus on strong interaction situations, such as merging scenarios, the method exhibits transferability across various strong interaction settings, like T-intersections and left turns at crossroads, as shown in **Fig. 24**. Theoretically, the proposed method can account for the uncertain behavior of other traffic participants in these scenarios, enabling safe interaction decisions for AVs.

**2) Multi-agent interactions.** Furthermore, the proposed method will be extended to processes involving multi-agent interactions, as shown in **Fig. 24**(c), (d). Given that it mathematically models the interaction between two strongly interacting agents, it can address multi-vehicle interaction scenarios by dividing multiple interacting vehicles into pairs. By expanding the interaction object matrix, this method can output optimal decision strategies beneficial for the overall traffic environment.

**3) Diverse driving styles.** The experimental process is benchmarked against the interactive behavior and driving trajectories of human drivers. Compare with diverse drivers, the proposed method achieves a high degree of average human-like decision-making similarity in naturalistic driving datasets of 188 test sequences (81.73%) and 145 dynamic interactions sequences (77.12%). Additionally, the method also demonstrates high similarity with behavioral decision-making of experienced and diverse drivers in real-vehicle tests. Therefore, it can be deduced that the method exhibits a certain level of adaptability to different driving styles and decision-making strategies of drivers.

Overall, the adaptive behavioral decision-making method demonstrates considerable adaptability and universality across various interactive scenarios, multiple traffic participants and diverse driving styles.

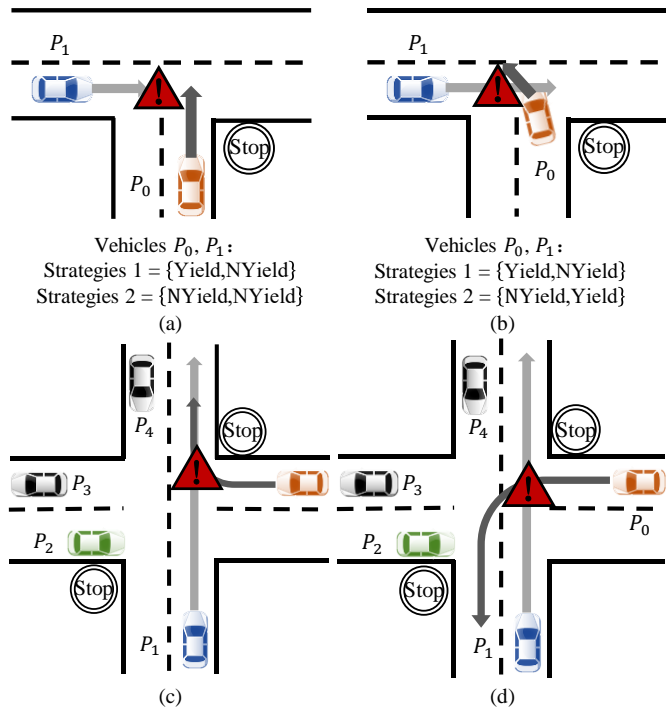

**Fig. 24.** The detailed interaction process for different driving cases. (a) Case 1 (b) Case 2. (c) Case 3. (d) Case 4

## VII. Conclusion

In this paper, we proposed an adaptive behavioral decision making approach for AVs, focusing on a ramp merging case study where it is important to consider the uncertain and strong interactions among vehicles. We first framed the problem based on non-cooperative game theory and derived the vehicle interaction behavior model. Based on the behavior probability output, the expected characteristics of the interactive behavior are obtained. Then, a model parameter optimization method based on maximum entropy IRL, is proposed to identify the optimal matching parameters in multiple environments through offline learning. Finally, through a mapping model between environmental variables and model parameters, AVs achieve to leverage dynamic traffic information, and learn online to obtain optimal parameters for the uncertain interactive vehicle behaviors. Moreover, experiments based on naturalistic driving dataset and real vehicle tests reveal that the proposed approach is robust enough to handle various situations, and its adaptability to the environment is highly consistent with the experience characteristics of actual driver interaction behavior. For highD and exiD datasets, the human-like decision-making similarity of the proposed approach is 81.73% in the test scenario and 77.12% in the dynamic random scenario. For real vehicle tests, the method demonstrates 72.73% human-like decision-making in tests, 69.84% in interactive scenarios, and maintains high safety by preventing safety violations, suggesting reduced collision risks. This adaptive ability to make decisions like humans in dynamic interaction scenarios is critical for the large-scale integration of AVs on public roads.

Future work will focus on improving the transferability of the algorithm in different strong interaction scenarios, such as overtaking on the highway, turning left at an intersection, etc. Further, the method can also be extended to model interaction processes involving the uncertain behavior of other traffic participants, including cyclists and pedestrians.

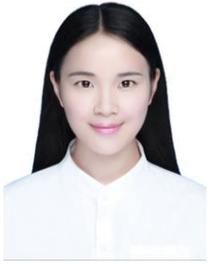

**Heye Huang** received the B.E. degree from Central South University, Changsha, China in 2018, and the Ph.D. degree from Tsinghua University, Beijing, China in 2023. She was invited as a Visiting Scholar at the Department of Cognitive Robotics, Delft University of Technology, from 2021 to 2022. She is currently a Research Associate Fellow with Connected & Autonomous Transportation Systems Lab, University of Wisconsin–Madison. Her current research interests include connected and automated vehicles, risk assessment, decision making and motion prediction.

Dr. Huang was a recipient of Outstanding Ph.D. Graduate, Outstanding Academic Star, the National Scholarship, and the First-Class Scholarship at Tsinghua University. She received the Best Paper Award at the International Symposium on Accident Analysis & Prevention in 2021, the Journal Cover Paper Award for Engineering in 2021, and the Best Research Award for Risk and Uncertainty in Engineering Systems in 2022.

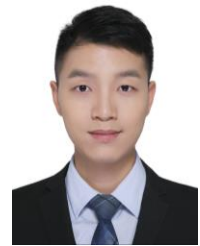

**Jinxin Liu** received the B. Tech. degree from Hefei University of Technology, Hefei, Anhui, China, in 2017 and the Ph.D. degree from Tsinghua University, Beijing, China in 2022. His research interests include vehicle trajectory prediction and behavioral decision making for autonomous vehicles.

Dr. Liu was a recipient of Outstanding Ph.D. Graduate, Excellent Doctoral Dissertation Award, and the Second-Class Scholarship at Tsinghua University. He received the Best Paper Award at the International Symposium on Accident Analysis & Prevention in 2021.

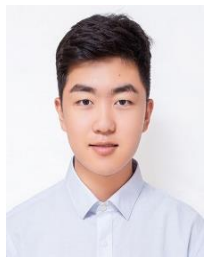

**Shiyue Zhao** (Student Member, IEEE) receives the B.S. degree in traffic engineering from Central South University, Hunan, China, in 2021. He is currently pursuing the Ph.D. degree with School of vehicle and mobility, Tsinghua University, Beijing, China. His research interests include intelligent transportation systems, advanced vehicle control, vehicle extreme control.

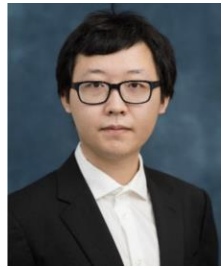

**Boqi Li** received the B.S. degree in mechanical engineering from the University of Illinois Urbana–Champaign, the M.S. degree in mechanical engineering from Stanford University, and the Ph.D. degree in mechanical engineering from the University of Michigan, Ann Arbor, in 2022. His research focuses on the modeling and decision-making for multiagent systems, particularly for the application of connected, cooperative, and automated mobility systems, where the navigation and motion planning of connected and automated vehicles are involved.

His research interests include human driver behavior prediction with machine learning, multi-agent reinforcement learning, and graph neural networks.

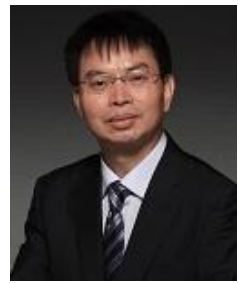

**Jianqiang Wang** received the B. Tech. and M.S. degrees from Jilin University of Technology, Changchun, China, in 1994 and 1997, respectively, and the Ph.D. degree from Jilin University, Changchun, in 2002. He is currently a Professor with the School of Vehicle and Mobility, Tsinghua University, Beijing, China.

He has authored over 180 papers and is a co-inventor of over 150 patent applications. His research interests include intelligent vehicles, driving assistance systems, and driver behavior. He was a recipient of the Best Paper Award in the 2014 IEEE Intelligent Vehicle Symposium, the Best Paper Award in the 14th ITS Asia Pacific Forum, the Best Paper Award in the 2017 IEEE Intelligent Vehicle Symposium. He was also a recipient of the "Changjiang Scholar Program Professor" in 2017, the Distinguished Young Scientists of NSF China in 2016, and the New Century Excellent Talents in 2008.